\documentstyle [11pt, epsfig]{article}
\begin{document}
 \def\la{\langle}
 \def\ra{\rangle}
\hfill {WM-05-116}

\vskip 1in   \baselineskip 24pt

{
\Large
   \bigskip
   \centerline{Kaluza-Klein Mesons in Universal Extra Dimensions}
 }
\def\bar{\overline}

\centerline{Erin De Pree\footnote{Email: ekdepr@wm.edu} and Marc 
Sher\footnote{Email: sher@physics.wm.edu}}
\centerline {\it Particle Theory Group}
\centerline {\it Department of Physics}
\centerline {\it College of William and Mary, Williamsburg, VA 23187, 
USA}
\bigskip

{\narrower\narrower In models with universal extra dimensions, the 
isosinglet Kaluza-Klein (KK) quarks ($q^{1}$)
have very narrow widths, of $O(5-10)$ MeV, and will 
thus hadronize.  Studies of KK-quarkonia ($\overline{q}^{1}q^{1}$) 
show very sharp resonances 
and dramatic signatures at the Linear Collider.  In this Brief Report, 
we consider the possibility of detecting KK-mesons, 
$(\overline{q}^{1}q$), and show that detection at a Linear 
Collider is unlikely.}

\vskip 1.0cm

In order for hadronic bound states to form, the constituents must 
have lifetimes longer than the hadronization time scale.   When the 
top quark was discovered to have a mass well in excess of $130$ GeV, 
it became clear that hadrons containing top quarks could not exist.  
In models beyond the standard model, strongly interacting states with 
sufficiently long lifetimes can certainly exist.  For example, a 
fourth generation quark, with very small mixings with lighter 
generations, could exist, and their bound states have been 
studied \cite{fourth}.  In supersymmetric (SUSY) models in which the 
gravitino is the lightest SUSY particle and in which a squark is 
the next-to-lightest, squarkonium \cite{squarkonium} and 
mesino \cite{mesino} bound states have been studied.

Recently, Carone et al. \cite{carone} considered bound states in models 
with universal extra dimensions \cite{ued}.  In these models, all states 
propagate in the higher dimensional space, and the existence of a 
Kaluza-Klein (KK) parity makes the lightest KK-state (LKP) stable.  It also 
allows for the compactification scale to be remarkably low, as low 
as $300$ GeV.  The 
other KK-states have masses only slightly greater than this lightest 
state, and they are thus long-lived.   Carone et al. analyzed bound 
states of KK-quarks, or KK-quarkonia.   In particular, the isosinglet 
KK-quarks will decay into a monochromatic quark and missing energy, 
leading to dramatic resonances, reminiscent of the $J/\psi$ and 
$\Upsilon$ states, with very clear signatures.  In this Brief Report, 
we study the possibility of detecting KK-mesons, consisting of a 
KK-quark and a zero-mode antiquark (or vice-versa).

In universal extra dimensions, the masses of the lightest excitations of 
the quarks, $q^{1}$, are degenerate with most of the other KK-states at tree 
level.  Radiative corrections \cite{cms} will break this degeneracy, 
leading to the KK-quarks being roughly $50-100$ GeV heavier than the 
LKP.  They calculated the widths for the KK-quarks
and found that the isosinglet $d^{1},s^{1}$ 
and $b^{1}$ KK-quarks were $O(5-10)$ MeV, for the decay into a quark and 
the LKP (which is mostly the KK-photon).  The widths of the $Q=2/3$
isosinglet KK-quarks are four times larger than those of the $Q=-1/3$ 
KK-quarks and will not be discussed 
further.

In Ref. \cite{carone}, it was claimed that the isosinglet KK-top quark was very long-lived 
(with a width of tens of keV), but they neglected mixing between the 
isosinglet and isodoublet KK-quarks.  The mass matrix for the KK-top 
modes is \cite{cms}
\begin{equation}
\pmatrix{ 1/R + \delta m_{T^{1}}&m_{top}\cr m_{top} & -1/R - \delta 
m_{t^{1}}\cr}
\end{equation}
where the $\delta m$'s are small radiative corrections.  This leads to a mixing 
angle which is given by $\tan 2\theta_{1}= 
2m_{top}R/(2+\delta m_{T^{1}}R+\delta m_{t^{1}}R)$.  This 
factor leads to an isosinglet top quark coupling to the 
b-quark and the KK-W boson given by the usual coupling times 
$\sin\theta_{1}$, and thus allow the KK-top to decay into those states 
directly.   We find the lifetime to be (assuming $|V_{tb}|=1$) given by
\begin{equation}
    \Gamma=\sin^{2}\theta_{1} {G_{F}\over\sqrt{2}}{M_{W}^{2}\over 3\pi m_{t^{1}}^{3}M_{W^{1}}^{2}}(m_{t^{1}}^{2}-M_{W^{1}}^{2})(m_{t^{1}}^{2}+2M_{W^{1}}^{2}).
\end{equation}
For $1/R\sim 500$ GeV, this is $10$ MeV.  As shown by in Ref.\cite{carone}, the signature will be a monochromatic b-quark, a 
monochromatic lepton and missing energy.

Given this width, hadronization will occur.  How could one detect 
the KK-mesons?  Recall how $B$ mesons are detected.  There are three 
signatures.   First, the 
$\Upsilon(4s)$ resonance is just above the threshold for a pair of $B$ 
mesons, and thus the strong decay causes the $\Upsilon(4s)$ to be 
much, much broader than the three lighter $\Upsilon$ states.  Second,
well above threshold, one can look at the $B$ meson decay products.  
Third, one can look for $B-\overline{B}$ mixing, and like sign dileptons.
We now examine each of these in turn.

One can produce copious numbers of $\overline{q}^{1}q^{1}$ mesons at a 
linear collider on resonance.  Above the threshold energy for the 
$\overline{q}^{1}q^{1}$
to decay to a $\overline{q}^{1}q$ meson and its antiparticle, the 
widths of the resonances become much larger.  In the WKB approximation, the number of 
states below threshold \cite{barger} is approximately 
$2\sqrt{m_{q^{1}}/m_{J/\Psi}}$, which gives $2$ for the $J/\Psi$ 
system, $3$ for the $\Upsilon$ 
system, and approximately 12 for the KK-quarkonium system.  As a 
result, one must look at the $13s$ state of KK-quarkonium.  However, 
Carone et al.\cite{carone} show that the production cross-section 
scales as $1/n^{3}$ \cite{carone}, and thus only the first 3 
resonances can be detected clearly.  As a 
result, this method of producing KK-mesons fails.

If one goes above threshold, could one detect the mesons through their 
decays?  Recall that the KK-quark will decay into a large amount of 
missing energy (typically 80 to 90 percent of the mass) plus a soft, 
monochromatic quark.  Given 
beamstrahlung (expected \cite{tesla} to be several GeV) and the 
expected beam resolution of at least 50 MeV, plus the huge amount of missing 
energy, it is hard to see how one could distinguish between a free 
KK-quark decay and one decaying in a meson.

Alternatively, one can look at the KK-meson decay, rather than the 
spectator quark decay.  Ignoring CKM angles, the KK-meson can 
annihilate through a KK-W into a KK-electron plus a 
neutrino, and the KK-electron then decays into a KK-photon plus an 
electron.   The width is given by
\begin{equation}
    {f^{2}_{M}g^{4}\over 64\pi  m_{q^{1}}^{3}}{m_{e^{1}}^{2}(m_{q^{1}}^{2}-m_{e^{1}}^{2})^{2}\over (m_{q^{1}}^{2}-m_{W^{1}}^{2})^{2}}
    \end{equation}
where $f_{M}$ is the meson decay constant, which is 
$O(\Lambda_{QCD})$.  Numerically, this is $O(10^{-7})$ GeV, which is 
negligible compared with the free  KK-quark width of a few MeV.  One 
can also consider the ``electromagnetic'' decay of a flavor-neutral 
KK-meson into a KK-photon plus a photon (analogous to 
$\pi^{o}\rightarrow \gamma\gamma$).   We have calculated this width, 
and find it to be approximately $5\times 10^{-6}$ GeV, which is also negligible.
None of this is 
surprising, since the decay constants give factors of 
$\Lambda_{QCD}^{2}$, which is much, much smaller than the other 
scales in the problem.

What about mixing?  In the case of the isosinglet KK-top quark, the 
mixing angles in the decay to a KK-W and a b-quark are large, and the 
sign of the lepton in the decay of the KK-W tags its charge, and thus 
the charge of the KK-top quark.  With mixing, one would see two like 
sign monochromatic leptons, which is a striking signature.  Sarid and 
Thomas \cite{mesino} showed that a mesino-antimesino oscillation, 
through this signature, could allow the discovery of mesinos, even if 
they couldn't otherwise be detected.   One must 
calculate the box diagram in which a W and a KK-W are exchanged.  For 
the KK-mesons $t^{1}\overline{q}$, we find the mass difference 
between the KK-meson and its antiparticle to be
\begin{equation}
\Delta m= 2\left({G_{F}\over\sqrt{2}}{\alpha\over 8\pi}\right)\left({1\over M_{W}\sin\theta_{W}}\right)^{2}\sum_{Q,Q^{1}}\Re \left[M^{4}_{W}V^{*}_{qQ}V_{t^{1}Q}V^{*}_{Q^{1}q}V_{Q^{1}t^{1}}{4\over 3}f^{2}_{t^{1}q}m_{t^{1}\overline{q}}A(m_{Q},m_{Q^{1}})\right]
    \end{equation}
where $Q$ is summed over $d,s,b$ and $Q^{1}$ is summed over 
$d^{1},s^{1},b^{1}$.  and
\begin{equation}
   A(m_{Q},m_{Q^{1}})= -\sum_{i=1}^{4}{M_{i}^{4}\ln M_{i}^{2}\over 
 \prod_{j\ne i}(M_{i}^{2}-M^{2}_{j})}
    \end{equation}
    where $M_{i} = (M_{Q},M_{W},M_{Q^{1}},M_{W^{1}})$.      
Alas, $\Delta m$ turns out to be utterly negligible, of the order of 
a few eV.    The reason is a 
 double-GIM mechanism---the $d,s,b$ quarks are nearly degenerate 
 (they are all very light compared to the other scales in the 
 problem), and the $d^{1},s^{1},b^{1}$ KK-quarks are also, in 
 universal extra dimensions, very nearly degenerate.  In the limit
 of exact degeneracy, the sum over the three generations will yield the 
 product of two columns in the CKM elements from the first two CKM 
 factors of Eq. (3), and the product of two rows from the latter two CKM 
 elements.  Thus, this 
 mechanism will also fail.
 
This is in sharp contrast with bound states of fourth generation 
quarks and supersymmetric quarks.  Fourth generation quarks can have 
longer lifetimes, neutral current (thus no missing energy) decays and 
the $Q=2/3$ quark will give a large GIM violation, leading to large 
mixing.  Supersymmetric quarks can also have longer lifetimes, less 
missing energy in their decays, and the mixing can occur through 
flavor-changing gluino interactions.  Thus, while bound states 
in those models are detectable, it appears as if 
Kaluza-Klein mesons are not.

We thank Chris Carone, Craig Dukes and Josh Erlich for many fruitful discussions.
This work was supported by the 
National Science Foundation grant PHY-023400.

\end{document}